\documentstyle[11pt]{article}
\bibliography{plain}
\title{\bf Remarks on the Cross Norm
Criterion for Separability } \vspace{20mm}
\author{S. J. Akhtarshenas  $^{a,b,c}$
\thanks{E-mail:akhtarshenas@tabrizu.ac.ir}
 , M. A. Jafarizadeh$^{a,b,c}$ \thanks{E-mail:jafarizadeh@tabrizu.ac.ir}
\\
\\
$^a${\small Department of Theoretical Physics and Astrophysics, Tabriz University, Tabriz 51664, Iran.} \\
$^b${\small Institute for Studies in Theoretical Physics and Mathematics, Tehran 19395-1795, Iran.} \\
$^c${\small Research Institute for Fundamental Sciences, Tabriz
51664, Iran.}} \pagebreak

 \newtheorem{thm}{Theorem}

 \newtheorem{prop}[thm]{Proposition}

\begin{document}
\maketitle \vspace{15mm}
\newpage
\begin{abstract}
Recently in Reference [ quant-ph/0202121] a computational
criterion of separability induced by greatest cross norm is
proposed by Rudolph. There, Rudolph conjectured that the new
criterion is not weaker than positive partial transpose criterion
for separability. We show that there exist counterexample to this
claim, that is,  proposed criterion is  weaker than the positive
partial transpose criterion.


{\bf PACs Index: 03.65.Ud }

\end{abstract}
\pagebreak

\vspace{7cm}

\section{Introduction}
Entanglement as the most non classical features of quantum
mechanics has been attracted much attention in the past decade.
Though, non local characters of quantum mechanics is singled out
in many decades ago \cite{EPR,shcro}, but it has recently received
considerable attention in connection with theory of quantum
information \cite{ben1,ben2,ben3}. Entanglement is usually arise
from quantum correlations between separated subsystems which can
not be created by local actions on each subsystems. By definition,
a bipartite mixed state $\rho$ is said to be separable if it can
be expressed as
\begin{equation}
\rho=\sum_{i}w_{i}\,\rho_i^{(1)}\otimes\rho_i^{(2)},\qquad w_i\geq
0, \quad \sum_{i}w_i=1,
\end{equation}
where $\rho_i^{(1)}$ and $\rho_i^{(2)}$ denote density matrices of
subsystems 1 and 2, respectively. Otherwise the state is
entangled.

The central tasks of quantum information theory is to characterize
and quantify entangled states. A first attempt in characterization
of entangled states has been made by Peres and Horodecki family
\cite{peres,horo}. It was shown that a necessary condition for
separability of a two partite system is that its partial transpose
be positive. Horodecki family  show that this condition is
sufficient for separability of composite systems only for
dimensions $2\otimes 2$ and $2 \otimes 3$.

A new criterion for separability and also an entanglement measure
for two partite systems based on greatest cross norm are
introduced by Rudolph \cite{rud1,rud2,rud3}. In an interesting
paper \cite{rud3} he obtained the values of greatest cross norm
for some states such as Werner states and isotropic states. In
\cite{rud3} Rudolph also introduced a computational criterion for
separability of mixed states induced by greatest cross norm and he
could obtain the separability conditions for some states such as
Werner states, isotropic states and 2-qubit Bell diagonal states.
He showed  that the new criterion completely characterizes
separability properties of pure states, Bell decomposable states
and isotropic states in arbitrary dimension. He conjectured that
new criterion is neither weaker nor stronger than positive partial
transpose (PPT) criterion introduced by Peres and Horodeckis in
\cite{peres,horo}.

In this paper we introduce Bell decomposable   states of $2\otimes
3$ systems  and we show that there is state in this category that
is entangled in the sense of PPT criterion but it is  separable in
the sense of new criterion introduced in \cite{rud3}, that is the
new criterion is weaker than PPT criterion.

The paper is organized as follows. In section 2 we briefly review
greatest cross norm criterion for separability of two partite
systems. The criterion induced by greatest cross norm is also
reviewed. Bell decomposable states in $2\otimes3$ systems are
introduced in section 3 and also PPT conditions for separability
of these states is obtained. Finally we show that there exist
state that is entangled in the sense of PPT criterion but satisfy
new criterion for separability proposed by Rudolph.

\section{Trace class norm criterion for separability and
associated induced separability criterion } In this section we
briefly review greatest cross norm for separability of two partite
systems introduced by Rudolph in \cite{rud1} and also the induced
criterion introduced in \cite{rud3}.

Let us consider Hilbert spaces $H_{1}$ and $H_{2}$ associated with
particles 1 and 2, respectively. One can show that the spaces
$T(H_1)$ and $T(H_2)$ of trace class operators on $H_1$ and $H_2$
are Banach spaces once they equipped with the trace class norm
$\|.\|_{1}^{(1)}$ and $\|.\|_{1}^{(2)}$, respectively. The
algebraic tensor product $T(H_1)\otimes_{alg}T(H_2)$ of $T(H_1)$
and $T(H_2)$ is defined as the set of all finite sums
$\sum_{i=1}^{n}u_i\otimes v_i$ where $u_i\in T(H_1)$ and $v_i\in
T(H_2)$ for all $i$.

A cross norm on $T(H_1)\otimes_{alg}T(H_2)$ is defined by (see
\cite{rud1,rud3} and references therein)
\begin{equation}
\|t\|_{\gamma}:=\inf\left\{\sum_{i=1}^{n}\|u_i\|_1\|v_i\|_1 \mid
t= \sum_{i=1}^{n}u_i\otimes v_i \right\},
\end{equation}
where $t\in T(H_1)\otimes_{alg}T(H_2)$ and the infimum is taken
over all finite decompositions of $t$ into elementary tensors. The
norm majorizes any subcross on $T(H_1)\otimes_{alg}T(H_2)$ (a norm
on $T(H_1)\otimes_{alg}T(H_2)$ is subcross norm if $\|t_1\otimes
t_2|\|\le \|t_1\|_1 \|t_2\|_1$ for all $t_1\in T(H_1)$ and $t_2\in
T(H_2)$ and it is cross norm if saturates the inequality for all
$t_1\in T(H_1)$ and $t_2\in T(H_2)$) is called greatest cross
norm.

The greatest cross norm criterion proposed by Rudolph is defined
as follows \cite{rud1,rud3}. Let $H_1$ and $H_2$ be finite
dimensional Hilbert spaces and $\rho$ be a density operator on
$H_1\otimes H_2$. The density matrix $\rho$ is separable if and
only if $\|\rho\|_{\gamma}=1$. Rudolph in \cite{rud3} determines
greatest cross norm for some states such as Werner states and
isotropic states. In addition in the second part of Ref.
\cite{rud3}, Rudolph introduced a new necessary separability
criterion for bipartite systems induced by the greatest cross norm
on Hilbert-Schmidt space.

In the Hilbert-Schmidt space $HS(H_1\otimes H_2)$, the operators
of the Hilbert space $H_1\otimes H_2$ are regarded as vectors.
This space is equipped with the Hilbert-Schmidt inner product
defined by $\left<T|T^{\prime}\right>=tr(T^{\dag}T^{\prime})$,
where $T$ and $T^\prime$ are two operators acting on space
$H_1\otimes H_2$. Let $HS(H_1)$ and $HS(H_2)$ denote
Hilbert-Schmidt spaces corresponding to Hilbert spaces $H_1$ and
$H_2$, respectively. It has been shown in \cite{rud3} that there
exist a one-to-one correspondence between Hilbert-Schmidt
operators $T\in HS(H_1\otimes H_2)$ and Hilbert-Schmidt operators
${\cal U}(T): HS(H_1)\longrightarrow HS(H_1)$. Alternatively we
can define the trace class norm of ${\cal U}(T)$ denoted by ${\cal
T}({\cal U}(T))$.

Every state $T\in HS(H_1\otimes H_2)$ in the Hilbert-Schmidt space
can be written as \cite{rud3}
\begin{equation}
T=\sum_{i}\lambda_i\, E_i\otimes F_i
\end{equation}
where $\{\lambda_i\}_i$ are non-negative real numbers and
$\{E_i\}_i$ and $\{F_i\}_i$ are orthonormal   bases of $HS(H_1)$
and $HS(H_2)$ respectively \cite{rud3}. Also the trace class norm
of ${\cal U}(T)$ is equal to ${\cal T}({\cal
U}(T))=\sum_{i}\lambda_i$.

Rudolph in \cite{rud3} proposed its new criterion for separability
in  a proposition which is quoted below:
\begin{prop}\cite{rud3}
Let $H$ be a finite dimensional Hilbert space and $\rho\in
T(H\otimes H)$ be a density operator. If $\rho$ is separable then
\begin{equation}\label{prop}
{\cal T}({\cal U}(\rho))\leq 1.
\end{equation}
\end{prop}

Based on the above criterion, Rodulph obtained separability
conditions of some states such as Werner states, isotropic states
and 2-qubit Bell diagonal states. He conjectured that the new
criterion is neither weaker nor stronger than the
Peres-Horodeckies PPT criterion for separability. In the next
section we present state that violates positive partial transpose
criterion but satisfy the separability criterion given in Eq.
(\ref{prop}).
\section{Bell decomposable states of $2\otimes 3$ quantum systems} In this
section we review Bell decomposable states of $2\otimes 3$ quantum
systems. A Bell decomposable density matrix acting on $2\otimes 3$
Hibert state can be defined by
\begin{equation} \label{BDS}
\rho=\sum_{i=1}^{6}p_{i}\left|\psi_i\right>\left<\psi_i\right|,\quad\quad
0\leq p_i\leq 1,\quad \sum_{i=1}^{6}p_i=1,
\end{equation}
where $\left|\psi_i\right>$ are Bell states in $H^2\otimes
H^3\cong H^6$ Hibert space , defined by:

$$
\left|\psi_1\right>=
\frac{1}{\sqrt{2}}(\left|11\right>+\left|22\right>), \qquad
\left|\psi_2\right>=
\frac{1}{\sqrt{2}}(\left|11\right>-\left|22\right>),
$$
\begin{equation}\label{BS}
\left|\psi_3\right>=
\frac{1}{\sqrt{2}}(\left|12\right>+\left|23\right>), \qquad
\label{BS4} \left|\psi_4\right>=
\frac{1}{\sqrt{2}}(\left|12\right>-\left|23\right>),
\end{equation}
$$
\label{BS5} \left|\psi_5\right>=
\frac{1}{\sqrt{2}}(\left|13\right>+\left|21\right>), \qquad
\label{BS6} \left|\psi_6\right>=
\frac{1}{\sqrt{2}}(\left|13\right>-\left|21\right>).
$$
It is quite easy to see that the above states are orthogonal and
thus span the Hilbert space of $2\otimes3$ systems.

A necessary condition for separability is presented by Peres
\cite{peres}. He show that the matrix obtained from from  partial
transpose of separable state must be positive. Horodeckies
\cite{horo} have shown that Peres criterion provides sufficient
condition for separability only for composite quantum systems of
dimension $2\otimes2$ and $2\otimes3$. This implies that the state
given in Eq. (\ref{BDS}) is separable if and only if the following
inequalities satisfy
\begin{eqnarray}\label{ppt}
(p_1+p_2)(p_3+p_4)\ge(p_5-p_6)^2, \\
(p_3+p_4)(p_5+p_6)\ge(p_1-p_2)^2, \\
(p_5+p_6)(p_1+p_2)\ge(p_3-p_4)^2.
\end{eqnarray}

On the other hand expanding Eq. (\ref{BDS}) in terms of canonical
base $\left|i\right>\otimes\left|j\right>$ we get
\begin{equation}\label{rhocan}
\begin{array}{rl}
\rho= & \frac{1}{2}(p_1+p_2)\left|11\right>\left<11\right|
+(p_1-p_2)\left|11\right>\left<22\right|
+(p_1-p_2)\left|22\right>\left<11\right| \\
+ & (p_1+p_2)\left|22\right>\left<22\right|
+(p_3+p_4)\left|12\right>\left<12\right|
+(p_3-p_4)\left|12\right>\left<23\right|  \\
+ & (p_3-p_4)\left|23\right>\left<12\right|
+(p_3+p_4)\left|23\right>\left<23\right|
+(p_5+p_6)\left|13\right>\left<13\right| \\
+ & (p_5-p_6)\left|13\right>\left<21\right|
+(p_5-p_6)\left|21\right>\left<13\right|
+(p_5+p_6)\left|21\right>\left<21\right|.
\end{array}
\end{equation}
Alternatively if
$\left|E_{ij}\right>\equiv\left|i\right>\left<j\right|$ denotes
corresponding bases in the Hilbert-Schmidt space then we have

\begin{equation}\label{rhoE}
\begin{array}{rl}
{\cal U}(\rho)= &
\frac{1}{2}(p_1+p_2)\left|E_{11}\right>\left<E_{11}\right|
+(p_1-p_2)\left|E_{12}\right>\left<E_{12}\right|
+(p_1-p_2)\left|E_{21}\right>\left<E_{21}\right| \\ + &
(p_1+p_2)\left|E_{22}\right>\left<E_{22}\right|
+(p_3+p_4)\left|E_{11}\right>\left<E_{22}\right|
+(p_3-p_4)\left|E_{12}\right>\left<E_{23}\right| \\ + &
(p_3-p_4)\left|E_{21}\right>\left<E_{32}\right|
+(p_3+p_4)\left|E_{22}\right>\left<E_{33}\right|
+(p_5+p_6)\left|E_{11}\right>\left<E_{22}\right| \\ + &
(p_5-p_6)\left|E_{12}\right>\left<E_{31}\right|
+(p_5-p_6)\left|E_{21}\right>\left<E_{13}\right|
+(p_5+p_6)\left|E_{22}\right>\left<E_{11}\right|,
\end{array}
\end{equation}
where $\left|E_{ij}\right>$ is used to denote $E_{ij}$.  Now we
can easily evaluate the  eigenvalues of $4\times 4$ matrix ${\cal
U}(\rho){\cal U}^{\dag}(\rho)$ which yields
\begin{equation}
\lambda_1=\lambda_2=A,\qquad \lambda_3=B+C, \qquad\lambda_4=B-C,
\end{equation}
where
$$
A=\frac{1}{4}\left((p_1-p_2)^2+(p_3-p_4)^2+(p_5-p_6)^2 \right),
$$
\begin{equation}
B=\frac{1}{4}\left((p_1+p_2)^2+(p_3+p_4)^2+(p_5+p_6)^2\right),
\end{equation}
$$
C=\frac{1}{4}
\left((p_1+p_2)(p_3+p_4)+(p_3+p_4)(p_5+p_6)+(p_5+p_6)(p_1+p_2)\right).
$$
It is easy to see that all eigenvalues are non-negative. Now we
can easily determine the separability criterion given in Eq.
(\ref{prop}) as
\begin{equation}
\sum_{i=1}^{4}\sqrt{\lambda_i}=2\sqrt{A}+\sqrt{B+C}+\sqrt{B-C}\le
1 .
\end{equation}

In the rest of this section we shall present a counterexample to
the claim that the criterion given in Eq. (\ref{prop}) is not
weaker than PPT criterion for separability. Let us consider a Bell
decomposable state given by
\begin{equation}\label{contexam}
p_1=0.3,\quad p_2=0,\quad p_3=0.2,\quad p_4=0.1,\quad
p_5=0.4,\quad p_6=0.
\end{equation}
It is quite easy to see that the state given by Eq.
(\ref{contexam}) violates PPT criterion given in Eq. (\ref{ppt}),
so it is entangled state. On the other hand, it is separable in
the sense of criterion given in Eq. (\ref{prop}). This implies
that the new criterion induced by the greatest cross norm is
weaker than the PPT criterion for separability.

\section{Conclusion }
We have provided a counterexample to show that the newly proposed
criterion of separability (induced by greatest cross norm) is
proposed by Rudolph is  weaker than the positive partial transpose
criterion, therefore, we have still a long way ahead  to solve the
long standing separability criterion in mixed quantum states.


\begin{thebibliography}{99}
\bibitem{EPR}{\sc A. Einstein, B. Podolsky and Rosen, }
{\em  Phys. Rev. {\bf 47}, 777 (1935).}
\bibitem{shcro}{\sc E. Schr\"{O}dinger, }{\em Naturwissenschaften.
 {\bf 23} 807 (1935).}
\bibitem{ben1}{\sc C. H. Bennett, and S. J. Wiesner,}
 {\em Phys. Rev. Lett. {\bf 69}, 2881 (1992).}
 \bibitem{ben2}{\sc C. H. Bennett, G. Brassard,
  C. Cr\'{e}peau, R. jozsa, A Peres and W. K. Wootters,}
 {\em Phys. Rev. Lett. {\bf 70}, 1895 (1993).}
\bibitem{ben3}{\sc C. H. Bennett, D. P. DiVincenzo, J. A. Smolin and W.K.
Wootters,} {\em Phys. Rev. A {\bf 54}, 3824 (1996).}
\bibitem{peres}{\sc A. Peres, }
{\em Phys. Rev. Lett. {\bf 77} 1413 (1996).}
\bibitem{horo}{\sc M. Horodecki, P. Horodecki and R. Horodecki, }
{\em Phys. Lett. A  {\bf 223} 1 (1996).}
\bibitem{rud1}{\sc O. Rudolph, }
{\em J. Phys. A {\bf 33} 3951 (2000).}
\bibitem{rud2}{\sc O. Rudolph, } {\em J. Math Phys. {\bf 42} 2507 (2001).}
\bibitem{rud3}{\sc O. Rudolph, }
 { \em eprint quant-ph/0202121. {\bf }}
\end{thebibliography}
\end{document}